\documentclass[twocolumn, prb]{revtex4}
\usepackage[ps2pdf,bookmarks=false,urlcolor=red,citecolor=blue,linkcolor=blue,
               pagecolor=red,hyperindex,colorlinks,hyperfigures
          ]{hyperref}
\usepackage[a4paper,left=1.8cm,right=1.8cm,top=3cm,bottom=2cm]{geometry}
\usepackage{amsmath, amssymb, amsthm}
\usepackage{graphicx}
\usepackage{times}

\begin{document}

\title{Whispering gallery mode enhanced optical force with resonant
tunneling excitation in the Kretschmann geometry}

\author{J. J. Xiao,$^{1,}$\footnote{Electronic mail: jjxiao@ust.hk} Jack
Ng,$^{1,}$\footnote{Electronic mail: jack@ust.hk} Z. F. Lin,$^{2}$
and C. T. Chan$^{1}$ }

\affiliation{$^1$Department of Physics and William Mong Institute of
NanoScience and Technology, The Hong Kong University of Science and
Technology, Clear Water Bay, Hong Kong, China
\\$^2$Surface Physics Laboratory and Department of Physics, Fudan
University, Shanghai 200433, China}

\begin{abstract}
The boundary element method is applied to investigate the optical
forces when whispering gallery modes (WGMs) are excited by a total
internally reflected wave. Such evanescent wave is particularly
effective in exciting the high-$Q$ WGM, while the low angular or
high radial order modes are suppressed relatively. This results in a
large contrast between the forces on and off resonance, and thus
allows for high size-selectivity. We fully incorporate the
prism-particle interaction and found that the optical force behaves
differently at different separations. Optimal separation is found
which corresponds to a compromise between intensity and $Q$ factor.
\end{abstract}
\date{\today}
 \maketitle

\newpage

Microcavities with whispering gallery mode (WGM) are important
components in nanophotonics, nonlinear optics, quantum optics, and
many other areas.~\cite{Optical:2004} As it is very difficult to
accurately control the resonance frequency or the size of a
micro-cavity during its fabrication, a convenient way to pick up the
particles with the desired resonance frequencies or size is
highly desirable. Recently, it was proposed in Ref.~2 
that the optical force acting on a microsphere under an evanescent
wave illumination~\cite{Kawata:1992, Chang:1997,Lester:1999,
Haumonte:2004, Jaising:2004} is highly size-selective: at resonance
the microsphere experiences a sizable force, whereas at off
resonance the force is negligible.~\cite{Other:2007} Utilizing this
property and the high quality factor ($Q>10^7$) of
WGM,~\cite{Optical:2004, Hara:2003} the size-selective force would
allow for accurate size-sorting of microsphere cavities, with an
accuracy of $\sim 1/Q$. However, the generation of evanescent wave
requires a substrate. Ref.~2 considers only the regime where the
microsphere is away from the near field of the substrate, and the
interaction of the microsphere and substrate is only treated in a
phenomenological manner.

\begin{figure}[b]
\centerline{\includegraphics[width=0.45\textwidth,clip]{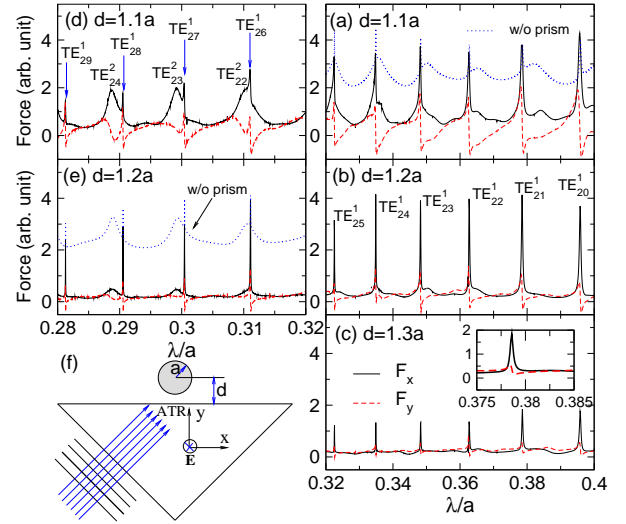}}
\caption{(Color online) Optical forces at different height from the
prism as functions of the wavelength. (Blue) dotted line: a
micro-cylinder illuminated by a plane wave in the absence of the
prism. (Black) solid line: the horizontal force $F_x$ acting on a
micro-cylinder with the prism. (Red) dashed line:  the vertical
force $F_y$ with the prism. (a) $d=1.1a$, (b) $d=1.2a$, (c)
$d=1.3a$, Inset of (c): an enlarged section of (c). (d) $d=1.1a$ and
(e) $d=1.2a$ for shorter wavelength. The $Q_{\text {iso}}$ and
resonant wavelength for the typical modes in (d) and (e) are,
respectively, $5399$ and $0.31110a$ for ${\text {TE}}_{26}^1$,
$7632$ and $0.30047a$ for ${\text {TE}}_{27}^1$, $10817$ and
$0.29055a$ for ${\text {TE}}_{28}^1$, $15368$ and $0.28128a$ for
${\text {TE}}_{29}^1$, $96$ and $0.31044a$ for ${\text
{TE}}_{22}^2$, $115$ and $0.29928a$ for ${\text {TE}}_{23}^2$, $139$
and $0.28892a$ for ${\text {TE}}_{24}^2$, and $170$ and $0.27926a$
for ${\text {TE}}_{25}^2$. (f) Illustration of the Kretschmann
geometry, where the evanescent wave is generated by total internal
reflection on top of the prism.} \label{fig1}
\end{figure}

In this article, using the boundary element method
(BEM)~\cite{Xiao:2008} and the Maxwell stress
tensor,~\cite{Lin:2005} we intend to go beyond Ref.~2 by fully
incorporating the influence of the substrate. We confine ourselves
to 2D calculations with a transverse electric (TE) polarization,
since 3D calculation is difficult due to the need of extremely fine
meshes in the BEM calculation. We expect that our result should be
qualitatively relevant for the 3D system. We consider the
Kretschmann geometry as depicted in Fig.~\ref{fig1}(f): a
micro-cylinder (MC) of radius $a$ sits on top of a prism, both have
a refractive index of $n_{\text p}=1.5$. The combined system (in
air) is then illuminated by an incident plane wave propagating at
$\theta=45^\circ$. As the critical angle of the prism is $\theta
_{\text C}=41.8^\circ$, the incident wave is total internally
reflected at the prism surface, and an evanescent wave is generated
there. The transfer of photon momentum to the MC is allowed by the
scattering or tunneling of the evanescent wave. We shall see that
when the particle is not in the near field of the prism surface, the
treatment of Ref.~2 gives a correct description of the situation.
For the near field case, we observe a series of interesting
phenomena, some are caused by the interplay between the increasing
intensity and the decreasing $Q$ as the particle approaches the
surface, and some are caused by the interaction between the particle
and the substrate.

The research on WGM-induced forces~\cite{Chylek:1978,
Almaas:1995,Arias:2000, Carmon:2005, Chan:2005, Eichenfield:2007,
Stilgoe:2008} can be dated back to the early eighties when Ashkin
reported the first observation of the force in
air.~\cite{Ashkin:1980} Later, the WGM-induced forces were also
observed by Fontes \textit{et al} in water.~\cite{Fontes:2005}
Nevertheless, only a modest enhancement in the optical force has
been achieved owing to the use of propagating waves in these works.
Meanwhile, it is well-known that the Kretschmann geometry can
efficiently excite the WGM of a high-$Q$ microsphere. Kawata and
Sugiura~\cite{Kawata:1992} demonstrated versatile optical
manipulation by evanescent waves subsequently.\cite{Dholakia:2008}
Although the Kretschmann geometry is well-known to be suitable for
both optical manipulation and the excitation of WGM, the resonant
forces associated with evanescent wave excited WGM have never been
considered experimentally. Here, we hope to provide a more complete
treatment for the WGM induced forces, and hope our work can
stimulate experimental works in this area.

In Figs.~\ref{fig1}(a)--\ref{fig1}(c), we show the optical forces
acting on a MC for $\lambda=0.32a-0.4a$ (size parameter $k_{\text
p}a\approx24-30$). The (blue) dotted line represents both the
horizontal and vertical forces (equal by symmetry) acting on a bare
MC without the prism.\cite{Our:2007} The (black) solid line
represents the horizontal force $F_{x}$ and the (red) dashed line
represents the vertical force $F_{y}$, in the presence of the prism.
For the case without the prism, the plane incident wave hits
directly on the MC, and for the case with the prism, the plane wave
is partially reflected by the prism surfaces and partially converted
into an evanescent wave on the prism's top surface. In the case of a
bare MC, it can be seen from Fig.~\ref{fig1}(a) that the force at
resonance is comparable to the force at off resonance; thus
size-selectivity is weak. On the other hand, the case with an
incident evanescent wave (generated by the prism) is very
size-selective, as we shall see below.

Let us first consider the horizontal force $F_{x}$. As shown in
Fig.~\ref{fig1}(c) where the separation $d=1.3a$, it is quite clear
that the force on first radial order resonance is significantly
stronger than the force at off resonance, and force on the second
order resonances can hardly be seen at all. This results in a high
peak-to-baseline ratio, defined as the peak resonance force divided
by the maximum of off resonance or lower-$Q$ resonance force. This
would imply the possibility of size-selective manipulation: upon
illumination, only those particles whose sizes happen to be in
resonance with the incident wave will experience a sizable optical
force, whereas the off resonance particle will be left untouched. In
the limit of large separation, the peak-to-baseline ratio can be
several orders of magnitude higher than what is predicted here, as
found out in Ref.~2. Indeed, the results presented in
Fig.~\ref{fig1}(c) already begin to resemble that of Ref.~2.

Although it is highly desirable to have a high peak-to-baseline
ratio, it is also important to achieve large force. This case is
illustrated in Figs.~\ref{fig1}(a) and \ref{fig1}(b). As the
separation decreases from $d=1.3a$ in Fig.~\ref{fig1}(c) to $d=1.2a$
in Fig.~\ref{fig1}(b), the force on first order resonance increases
by a factor of $\sim $2--3 as the intensity becomes higher in the
near field, while its linewidth is broadened slightly. At $d=1.2a$,
the second order resonance can now be observed, due to the higher
intensity. Although the peak-to-baseline ratio for $d=1.2a$ is
smaller than that of $d=1.3a$, the size selectivity is still very
strong. At an even closer separation [$d=1.1a$ in
Fig.~\ref{fig1}(a)], although the intensity of the evanescent wave
increases, the strength of the first order resonance force remains
at the same level as when $d=1.2a$. This is a consequence of the
Lorentzian lineshape,~\cite{and:2008} which implies that the
resonance force is inversely proportional to the linewidth, or
directly proportional to its $Q$. As $d$ decreases, the stronger
modal coupling between the prism and the MC broadens the linewidth
(or decrease in $Q)$, and causes the resonance force to decrease,
compensating the increase in the local intensity $I_{\text loc}$.
The background force and the forces associated with the second order
resonance have increased considerably, making the peak-to-baseline
ratio small. We note that as the MC approaches the prism, the
resonance frequencies do not shift and it is the $Q$'s that drops.
This decrease in $Q$ will significantly reduce the size-sensitivity
of the resonance force, which scales like $\sim $1/$Q$.

A more explicit demonstration on the selectivity due to high-$Q$ is
shown in Fig.~\ref{fig1}(d) and \ref{fig1}(e) which show the optical
forces with the presence of the prism in higher size parameter
regime $\lambda =0.28a-0.32a$ ($k_{\text p} a\approx 30-34$). It is
seen in Fig.~\ref{fig1}(d) that with $d=1.1a$ several radially
second order WGMs can be well excited, resulting in comparable
optical forces of the radially first order WGMs as indicated by the
downward arrows. We find that the $Q$ value of the ${\text
{TE}}_{23}^2$ mode exceeds $10^2$ (see the details in the caption of
Fig.~\ref{fig1}). The MC with such a value of $Q$ can already
collect sufficient incident field energy that yields a
non-negligible amount of light scattering. Figure~\ref{fig1}(e)
shows that for increased separation to $d=1.2a$, these second order
modes remain not well excited again, simply because of the decreased
local field. At this separation, however, the radially first order
WGMs with $Q$ in the order of $10^3$ and $10^4$ can be well excited
and are manifested as the several sharp peaks [as labeled in
Fig.~\ref{fig1}(e)].

\begin{figure}[t]
\centerline{\includegraphics[width=0.25\textwidth,clip]{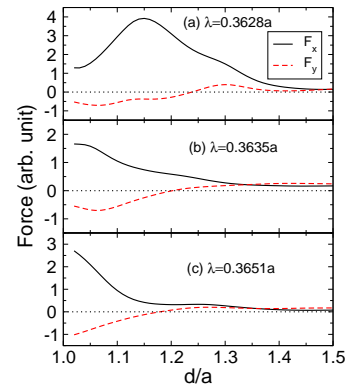}}
\caption{(Color online) Optical forces as a function of
cylinder-prism separation $d$ at various situations. (a) at the
${\text {TE}}_{22}^1 $ resonance, (b) at an off resonance frequency
between ${\text {TE}}_{22}^1$ and ${\text {TE}}_{18}^2 $, (c) at the
${\text {TE}}_{18}^2 $ resonance.} \label{fig2}
\end{figure}
Figure~\ref{fig2} clarifies the separation dependence of the optical
force on and off resonance. Here we have chosen three different
frequencies, one for ${\text {TE}}_{22}^1 $ (with quality factor of
isolated particle $Q_{\text {iso}}=1395$)~\cite{Xiao:2008} shown in
Fig.~\ref{fig2}(a), one for ${\text {TE}}_{18}^2 $ ($Q_{\text
{iso}}=49$)~\cite{Xiao:2008} shown in Fig.~\ref{fig2}(c), and the
last one for an off resonance frequency shown in Fig.~\ref{fig2}(b).
We remark that the strength of the resonance force depends on both
the local intensity $I_{\text{loc}}$ and the mode's $Q$, i.e.
$F\propto QI_{\text{loc}}$. The observed $Q$ in the force spectrum
is in fact related to the original quality factor for the isolated
MC, $Q_{\text {iso}}$, and the coupling $Q$-factor,
$Q_{\text{coup}}$, through the
equation$1/Q=1/Q_{\text{iso}}+1/Q_{\text{coup}} $. In
Fig.~\ref{fig2}(b) and \ref{fig2}(c), the optical force decreases
monotonically as $d$ increases. The off resonance case of
Fig.~\ref{fig2}(b) is straightforward to understand: as $d$
increases, the local intensity decreases, so the force decreases as
well. For the second order resonance${\text {TE}}_{18}^2 $ case
plotted in Fig.~\ref{fig2}(c), the system's $Q$ is completely
determined by its small $Q_{\text {iso}}=49$ (because $Q_{\text
{iso}}\ll Q_{\text {coup}})$, consequently as $d$ increases, the
total $Q$ is essentially unchanged while $I_{\text{loc}}$ decreases,
causing the optical force to drop. For the high-$Q$ resonance of
${\text {TE}}_{22}^1 $ shown in Fig.~\ref{fig2}(a), it can be
anticipated that the optical forces will be weak when the MC is
either too close to (due to a small $Q)$ or far away from (due to
small $I_{\text{loc}}$) the prism, and there is a peaked force in
between. In Fig.~\ref{fig2}(a), this peak occurs at $d\approx
1.16a$.

\begin{figure}[t]
\centerline{\includegraphics[width=0.35\textwidth,clip]{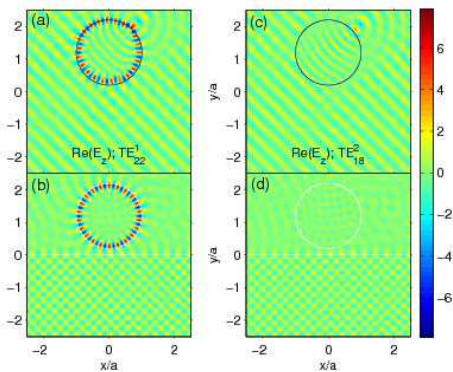}}
\caption{(Color online) Near fields for high-$Q$ (left column) and
low-$Q$ (right column) resonances. (a) and (c) are, respectively,
for ${\text {TE}}_{22}^1$ and  ${\text {TE}}_{18}^2$ in the absence
of the prism. (b) and (d) are, respectively, for the same resonances
as (a) and (c), but with the prism at a separation $d=1.2a$ (lower
panels).} \label{fig3}
\end{figure}

To give an overall perspective of the effect of $Q$ value and the
effect of the prism, we compare the near fields in Fig.~\ref{fig3}
for four different situations corresponding to Figs.~\ref{fig2}(a)
and \ref{fig2}(c). These figures clearly demonstrate the resonant
tunneling mechanism which accounts for the optical forces. We can
draw a conclusion that a high $Q$ value is the prerequisite for the
resonant tunneling. Therefore in the scheme with the prism for
optical sorting operated at a specific frequency, high $Q$ particles
can be easily distinguished and selected.

We finally consider the vertical force $F_{y}$. As Rayleigh
dielectric objects tend to move to region of high intensity in order
to minimize the electromagnetic free energy, one may believe that
the vertical force is attractive relative to the prism. But as shown
in Fig.~\ref{fig1}(c) where $d=1.3a$, the vertical force versus
wavelength curve actually consists of a background repulsion which
lifts the entire curve upward plus a peak and dip pair at each
resonance position. As the MC approaches the prism [Figs.
~\ref{fig1}(a) and \ref{fig1}(b)], the lineshape of $F_{y}$ evolves
and deviates from that of the original lineshape in
Fig.~\ref{fig1}(c). We see that the resonance force has become
stronger, and $F_{y}$ can now be either attractive (at the dip) or
repulsive (at the peak). This resonance lineshape (the peak and dip
pair) can be understood as follows: when the frequency changes
across the resonance frequency, the WGM switches from in phase to
out of phase with the incident field, and this results in a change
of the sign in the force. We note that in addition to typical
resonance peak-and-dip, there is still a bias of the force to the
repulsive side. Heuristically, this repulsive bias can be traced to
the fact that the incident photon momentum is upward, so it should
not be surprising to see the photons pushing the MC up. We note that
in the 1992 experiment of Kawata and Sugiura,~\cite{Kawata:1992} a
repulsive force is also observed. This subtle effect warrants
further studies.

In summary, we have numerically investigated the WGM enhanced
optical force as excited by an evanescent wave. The high
peak-to-baseline ratio associated with evanescent wave excitation
makes it highly size-selective and may thus be useful for particle
sorting. The notion of Ref.~2 is confirmed even in the presence of a
substrate, and we have gone beyond Ref.~2 and show that the idea
even works in the substrate's near field. In general, a high $Q$
mode guarantees a stronger optical force, as it favors the resonant
tunneling of the evanescent light. As a consequence of the
competition between the increased coupling loss at small separation
and the decreased intensity at large separation, there exists an
optimal particle-prism separation in which the resonance force is
maximized. Accordingly, in an optical sorting experiment, it would
be highly desirable to have some other means to keep the particles
at an appropriate height, for if the MC is too close to the prism
surface, the linewidth broadening would reduce the sorting accuracy
as well as the peak-to-baseline ratio, but if the MC is too far
away, the force is too weak. Finally, we remark that the induced
optical force can further be enhanced by coating the substrate with
a metallic coating that supports surface plasmon,~\cite{Garces:2006}
or by coating the prism with a dielectric cavity
layer.~\cite{Reece:2006}

\noindent This work was supported by the Hong Kong RGC Grant
No.~600308. ZFL was also supported by NSFC (10774028), PCSIRT and
MOE of China (B06011). Computational resources were supported by the
Shun Hing Education and Charity Fund. JJX acknowledges Jeffrey C. W.
Lee's technical help on the MPI.

\end{document}